\documentclass[twocolumn,prl,showpacs]{revtex4-1}

\usepackage{dcolumn}
\usepackage{amsfonts}
\usepackage{amsmath}
\usepackage{amssymb}
\usepackage{bm}
\usepackage{epsfig}
\usepackage[usenames]{color}

\begin{document}

%\ifpdf \DeclareGraphicsExtensions{.jpg,.pdf,.tif} \else
%\DeclareGraphicsExtensions{.eps,.jpg} \fi

\newcommand{\brm}[1]{\bm{{\rm #1}}}
\newcommand{\tens}[1]{\underline{\underline{#1}}}
\newcommand{\xv}{\bm{{\rm x}}}
\newcommand{\Rv}{\bm{{\rm R}}}
\newcommand{\uv}{\bm{{\rm u}}}
\newcommand{\nv}{\bm{{\rm n}}}
\newcommand{\Nv}{\bm{{\rm N}}}
\newcommand{\ev}{\bm{{\rm e}}}
\newcommand{\Cv}{\bm{{\rm C}}}
\newcommand{\Qv}{\bm{{\rm Q}}}
\newcommand{\qv}{\bm{{\rm q}}}
\newcommand{\fv}{\bm{{\rm f}}}
\newcommand{\tv}{\bm{{\rm t}}}
\newcommand{\kv}{\bm{{\rm k}}}
\newcommand{\Dv}{\bm{{\rm D}}}
\newcommand{\av}{\bm{{\rm a}}}
\newcommand{\bv}{\bm{{\rm b}}}
\newcommand{\Gv}{\bm{{\rm G}}}
\newcommand{\Tv}{\bm{{\rm T}}}
\newcommand{\pv}{\bm{{\rm p}}}
\newcommand{\wn}{m}
\newcommand{\rv}{\bm{{\rm r}}}

\newcommand{\cvh}{\hat{\brm{c}}}
\newcommand{\evh}{\hat{\brm{e}}}
\newcommand{\nvh}{\hat{\brm{n}}}
\newcommand{\avh}{\hat{\brm{a}}}
\newcommand{\bvh}{\hat{\brm{b}}}

\newcommand{\Ochange}[1]{{\color{red}{#1}}}
\newcommand{\Ocomment}[1]{{\color{PineGreen}{#1}}}
\newcommand{\Tcomment}[1]{{\color{ProcessBlue}{#1}}}
\newcommand{\Tchange}[1]{{\color{BurntOrange}{#1}}}
\newcommand{\Remove}[1]{{}}

\title{Topological phonons and Weyl lines in 3 dimensions}

\author{Olaf Stenull}
\affiliation{Department of Physics and Astronomy, University of
Pennsylvania, Philadelphia, PA 19104, USA }

\author{C. L. Kane}
\affiliation{Department of Physics and Astronomy, University of
Pennsylvania, Philadelphia, PA 19104, USA }

\author{T. C. Lubensky}
\affiliation{Department of Physics and Astronomy, University of
Pennsylvania, Philadelphia, PA 19104, USA }

\vspace{10mm}
\date{\today}

\begin{abstract}
Topological mechanics and phononics have recently emerged as an
exciting field of study. Here we introduce and study
generalizations of the three-dimensional pyrochlore lattice
that have topologically protected edge states and Weyl lines in
their bulk phonon spectra, which lead to zero surface modes
that flip from one edge to the opposite as a function of
surface wavenumber.
\end{abstract}

\pacs{62.20.D-, 03.65.Vf}

\maketitle

\begin{figure}[ptb]
	%\centering{\includegraphics[width=4.2cm]{origPyro}}
	%\hspace{-0.5cm}
	%\centering{\includegraphics[width=4.2cm]{plusOnePyro}}
	%\centering{\includegraphics[width=4.2cm]{minusOnePyro}}
	%\hspace{0.2cm}
	%\centering{\includegraphics[width=3.5cm]{unitCellPlot}}
	%\centering{\includegraphics[width=4cm]{unitCellRedo}}
    \centering{\includegraphics{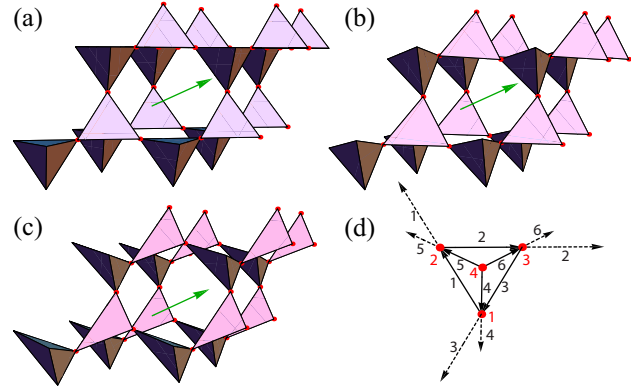}}
	\caption{(Color Online) (a) The original pyrochlore lattice
		and the GPL in
		the (b) $X_1$ and (c) $X_{-1}$ conformations (see text). The
		(green) arrow represents the $(1, 1, 1)$-direction. (d) The
		unit cell of (a). The (red) dots with the (red) numbers mark
		the basis sites. The solid (dashed) arrows designate the bond
		vectors $\brm{a}_b$ ($\brm{a}^\prime_b$) of the internal (external) bonds.}%
	\label{fig:origAndGenLattice}%
\end{figure}

Mechanical lattices with a perfect balance between the number
of degrees of freedom and the number of constraints (springs)
with unit cells of appropriate internal geometry exhibit
zero-frequency modes at boundaries even though they have very
few if any zero modes in their bulk
~\cite{LubenskySun2015,SunLub2012}. The topological origin of
these zero modes was explained in Ref.~\cite{KaneLub2014},
which introduced the framework for a phononic version of
topological band theory that is well known in electronic
contexts, including polyacetylene \cite{ssh}, quantum Hall
systems~\cite{halperin82,haldane88}, and topological
insulators~\cite{km05b,bhz06,mb07,fkm07,HasanKane2010,QiZhang2011}.
So far, attention has focused on one-~\cite{ChenVit2014} and
two-dimensional
(2$d$)~\cite{PauloseVit2015,PauloseVit2015-b,RocklinLub2016,SussmanLub2016}
model systems, such as the generalized kagome lattice (GKL)
introduced in Ref.~\cite{KaneLub2014}, which places masses and
central-force springs on a lattice of corner-sharing triangles.
In this paper, we consider the natural three-dimensional (3$d$)
generalization of this system -- a generalized pyrochlore
lattice (GPL) with deformed corner sharing tetrahedra, whose
edges are occupied by central-force springs, as depicted in
Fig.~\ref{fig:origAndGenLattice}.  We show that the bulk phonon
spectrum of these lattices exhibit lines of topologically
protected zero modes, analogous to lines of touching bands in
line-node semimetals
\cite{Volovik2007,BurkovBal2011,KimRap2015,YuHu2015} and gyroid
photonic crystals \cite{LuJoan2013}, that cause the number of
protected surface modes to undergo discontinuous jumps as a
function of surface wavenumber.

In 2$d$ GKLs, there exist classes of conformations in which all
of the bulk vibrational modes, except for the zero-wavevector
acoustic modes, have a finite frequency. These fully ``gapped"
systems fall into topological classes distinguished by
topological invariants characterizing their reciprocal-space
band structure. Topological zero-frequency edge modes arise
from an interplay between this bulk topological structure and
the local mismatch in the number of degrees of freedom and
constraints, so that given the bulk topological invariants and
the local structure of the surface termination, it is possible
to predict the number of zero-frequency modes localized at a
surface. In GKLs, the transition between topologically distinct
phases is marked by the existence of a line in the Brillouin
zone along which normal-mode frequencies vanishes. This line of
zero modes is a consequence of the existence of straight lines
of bonds, which we call straight filaments, that carry a state
of self stress (SSS) in which the bonds are under tension but
site forces vanish. The Maxwell-Calladine counting rule (see
below) ~\cite{Maxwell1864,Calladine1978} applied with periodic
boundary conditions, then guarantees corresponding zero modes.

%%%%%%%%%%%%%%%%%%%%%%%%%%%
%%%%%%%%%%%%%%%%%%%%%%%%%%%

Three-dimensional GPLs have a richer structure than
two-dimensional GKLs. We find that, like GKLs, they exhibit
gapless bulk modes associated with localized SSSs: Arrays of
parallel lines of SSSs (straight filaments) -- four in the
undistorted pyrochlore lattice aligned along the tetrahedral
edges -- have associated planes of zero modes in reciprocal
space aligned perpendicular to the array filaments, and flat
planes of self-stress have associated perpendicularly aligned
lines of zero modes in reciprocal space. As in the GKL, the
bulk zero modes on these planes (lines) can be lifted to
non-zero frequency by distorting the lattice to remove the
straight lines (flat planes) of bonds. However, a new
phenomenon, which did not occur in the GKL, arises. There are
lines of bulk zero modes in reciprocal space that are protected
by an integer topological invariant defined on a path that
encircles them. These are the analog of Weyl lines that exist
in electronic
\cite{Volovik2007,BurkovBal2011,KimRap2015,YuHu2015} and
photonic~\cite{LuJoan2013} systems. Analogous Weyl/Dirac points
\cite{WanVish2011,BurkovBal2011,XuHa2015} can also be present
in 2$d$, as in the electronic spectra of graphene
\cite{DivincenzoMele1984,CastroGeim2009,YoungKane2015} and
phonon spectra in deformed square
lattices~\cite{RocklinLub2016} and 2$d$ models of jammed
matter~\cite{SussmanLub2016}, and they appear to be generic in
large-unit-cell Maxwell lattices. We will show that the
presence of Weyl lines in the bulk also has important
consequences for the number and location of surface zero modes.

Our GPL is based on the usual pyrochlore lattice with $4$-site,
$12$-bond unit cells. We choose our standard reference unit
cell to have sites at the $4$ corners of a tetrahedron at the
basis positions $\brm{r}_1 =\frac{1}{2} (1, 1, 0)$, $\brm{r}_2
= \frac{1}{2} (0, 1, 1)$, $\brm{r}_3 = \frac{1}{2} (1, 0, 1)$,
and $\brm{r}_4 = (0, 0, 0)$  and  bonds as shown in
Fig.~\ref{fig:origAndGenLattice} (d). Note that this cell lacks
inversion symmetry, as do all other unit cells in GPLs, and, as
a result, it has a nonvanishing dipole moment when charges
$d=+3$ are placed at its sites and charges $-1$ at the center
of its bonds. The primitive translation vectors of the lattice
are $\brm{T}_1 = (1, 1, 0)$, $\brm{T}_2 = (0, 1, 1)$ and
$\brm{T}_3 = (1, 0, 1)$. There are six internal bonds, defined
by vectors $\brm{a}_1$ to $\brm{a}_6$ that connect the sites
within the unit cell, and six external bonds, defined by
vectors $\brm{a}^{\prime}_1$ to $\brm{a}^{\prime}_6$ that
connect sites of a given unit cell to sites of nearest-neighbor
unit cells. The triangular faces of lattice tetrahedra lie in
four sets of parallel planes and form kagome lattices in each.
One of the four sets has its layer normal parallel to the $(1,
1, 1)$-direction. In the GPL, this latter set of planes will
play a distinguished role, and we will refer to it as the GKL
planes.

We generalize the pyrochlore lattice by allowing the basis
sites to deviate from their original reference positions. In
the undeformed pyrochlore lattice, there are 6 sets of straight
filaments, built from the bonds on the 6 edges of the
tetrahedra, that carry SSSs. Following Ref.~\cite{KaneLub2014},
our generalization is designed to convert specific straight
filaments into ``zigzagged" ones that do not carry SSSs. In the
GPL, the positions of the basis sites are displaced relative to
those of the reference lattice, $\brm{r}_i \to \brm{r}_i +
\delta \brm{r}_i (X)$, where $X$ is a shorthand for a set of 4
parameters, $X=(x_1, x_2, x_3, z)$. We set the deviations $
\delta \brm{r}_i (X)$ to be
\begin{subequations}
\label{deformationDef}
\begin{align}
\delta \brm{r}_1 (X) & = x_1 \sqrt{3} \evh_1 - x_2 \avh_3\, ,
\\
\delta \brm{r}_2 (X) & = x_2 \sqrt{3} \evh_2 - x_3 \avh_1 \, ,
\\
\delta \brm{r}_3 (X) & = x_3 \sqrt{3} \evh_3 - x_1 \avh_2 \, ,
\\
\delta \brm{r}_4 (X) & = - z \nvh\, ,
\end{align}
\end{subequations}
where $\avh_j = \brm{a}_j/| \brm{a}_j |$ and $\evh_j =
(\brm{a}_j \times \nvh)/|\brm{a}_j \times \nvh|$ with $\nvh$
the unit vector in the $(1, 1, 1)$-direction. If $z \neq 0$ and
$x_1=x_2=x_3=0$, there are three sets of straight and three
sets of zigzagged filaments, the former lying in the GKL
planes. Nonzero $x_n$'s produce the same distortions of the
kagome lattices in the GKL planes as those in
Ref.~\onlinecite{KaneLub2014} with each $x_n$ converting the
straight filament parallel to $\avh_n$ to a zigzag one. We have
also studied more general versions of our model lattice in
which sites $1$ to $3$ can adopt positions outside the GKL
planes, thereby destroying their independent SSSs. For
simplicity, however, we will focus on the model lattices
described by Eq.~(\ref{deformationDef}). More specifically, we
will focus on the 2 lattice conformations corresponding to the
parameter settings $X = X_1 \equiv (0.1, 0.1, 0.1, 0.1)$ and $X
= X_{-1} \equiv (-0.1, 0.1, 0.1, 0.1)$, see
Fig.~\ref{fig:origAndGenLattice}. The GKL planes of these
lattices are equivalent to the GKL conformations depicted in
Fig.~2 (c) and (e) of Ref.~\cite{KaneLub2014}, respectively.

Any central-force elastic network consisting of periodically
repeated unit cells with $n$ sites and $n_B$ bonds is governed
by the generalized Calladine-Maxwell
theorem~\cite{Maxwell1864,Calladine1978,LubenskySun2015}
$n_0(\qv) - s(\qv)=d n- n_B$  at each wavevector $\qv$ in the
BZ. It relates the number $n_0(\qv)$ of zero modes, whose
displacements do not stretch bonds, and the number $s(\qv)$ of
SSSs, in which bonds under tension exert no net forces on
sites, to the invariant properties  $n$ and $n_B$ of the unit
cell. This important relation follows from the the properties
of the $n_B \times dn$ compatibility matrix $\brm{C}(\qv)$
relating bond displacements $\uv(\qv)$ to bond extensions
$\ev(\qv)$ via $\brm{C}(\qv) \uv(\qv) = \ev(\qv)$ and the
$dn\times n_B$ equilibrium matrix $\brm{Q}(\qv)=\Cv^\dag(\qv)$
relating bond tensions $\tv(\qv)$ to site forces $\fv(\qv)$ via
$\Qv(\qv) \tv(\qv) = \fv(\qv)$. Zero modes constitute the null
space of $\Cv(\qv)$ and SSSs the null space of $\Qv(\qv)$. When
all masses and spring constants are set to unity, as we do
here, the dynamical matrix governing the phonon spectrum is
simply $\Dv(\qv) =\Qv(\qv)\Cv(\qv)$. Under periodic boundary
conditions, the GPL satisfies $dn=n_B$, i.e., it is a Maxwell
lattice~\cite{LubenskySun2015} in which at each $\qv$ including
$\qv=0$, there is always one SSS for each zero mode. The
compatibility, equilibrium, and dynamical matrixes are all
$12\times 12$ matrixes (see the supplemental material for
details).

The elastic energy density can be expressed \cite{Kittel1971}
in terms of the six-dimensional vector of symmetric strains $U
= (u_{xx}, u_{yy}, u_{zz}, u_{xy}, u_{xz}, u_{yz})$ and the $6
\times 6$ Voigt matrix $\brm{K}$: $f = \frac{1}{2} U^T\cdot
\brm{K}\cdot U$. We calculate the $\brm{K}$ of the GPLs from
the normalized eigenvectors of the null space of
$\Qv(\qv=0)$~\cite{LubenskySun2015}. The original pyrochlore
lattice has $s(\qv)=6$ corresponding to its six sets of
straight filaments. This provides a sufficient number of SSSs
at $\qv=0$ to stabilize all six independent strains. For $X_1$
and $X_{-1}$, there are no straight filaments in the lattice,
but $s(\qv=0) = 3$ because there are $n_0(\qv=0)=3$ zero modes
corresponding to the rigid translations of the lattice. Thus,
$\brm{K}$ has three zero and three positive eigenvalues for
these lattice conformations. The former imply the existence of
three independent macroscopic deformations free of restoring
forces -- the Guest modes \cite{GuestHut2003} $\brm{U}^G$ of
the lattice. For $X_1$, the three Guest modes are proportional
to $(2, 0, 0, 0, 0, -1)$, $(0, 2, 0, 0, -1, 0)$ or $(0, 0, 2,
-1, 0, 0)$. Internally, these deformations \Remove{which
combine uniaxial stretch or compression with shear in the
perpendicular plane} are realized by rigid rotations of the
constituent tetrahedra, about $\nvh$, that change no bond
lengths. For $X_{-1}$, the Guest modes involve all components
of $\brm{U}$.

\begin{figure}[ptb]
\includegraphics[width=4.2cm]{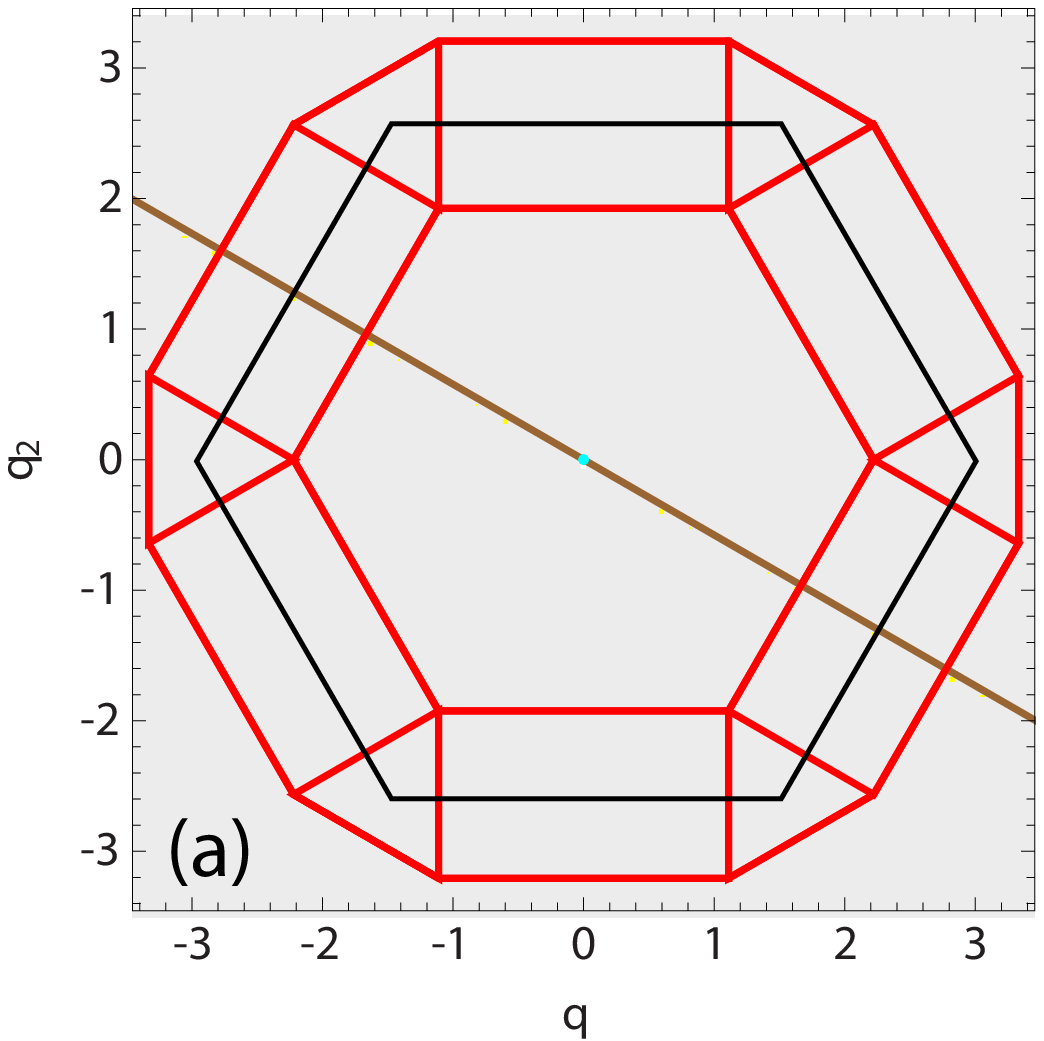}
\includegraphics[width=4.2cm]{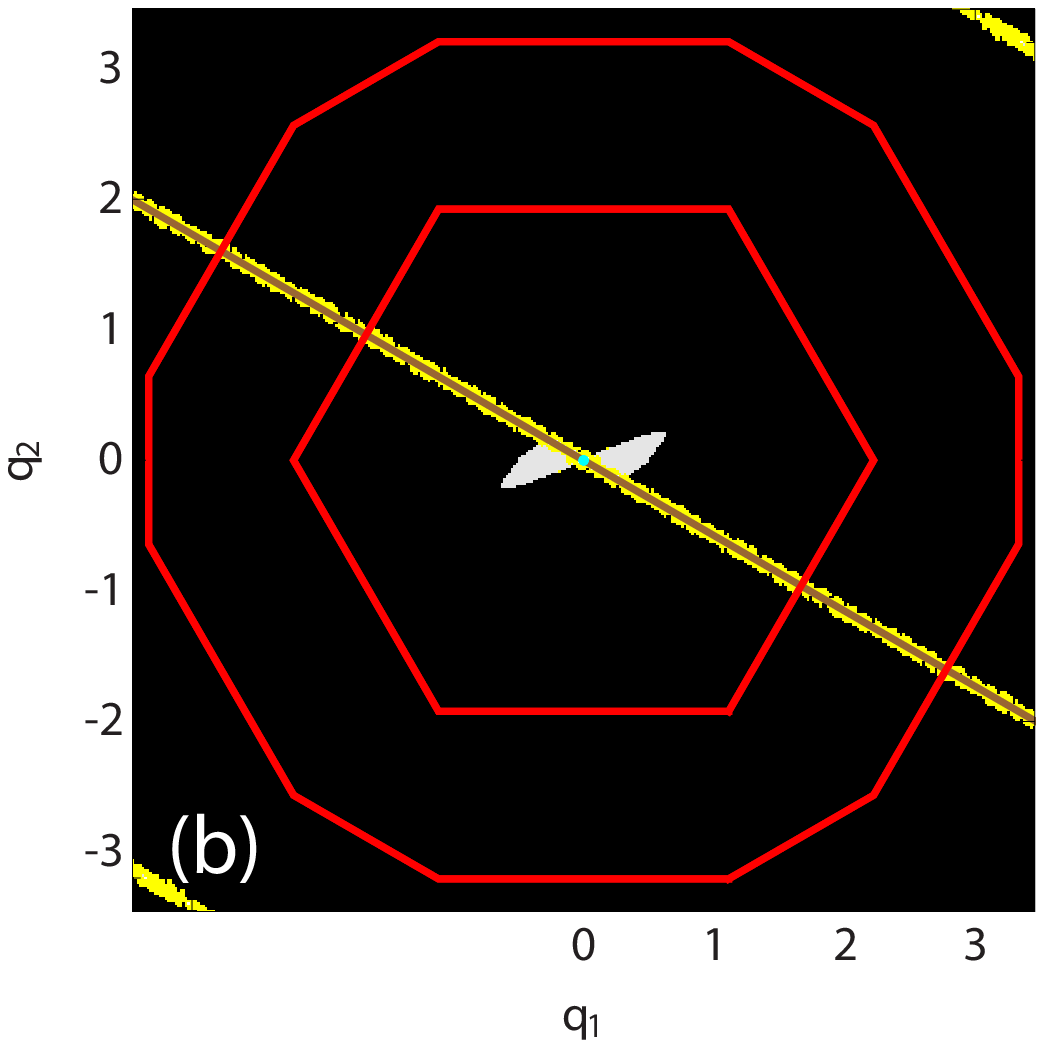}
\includegraphics[width=4.2cm]{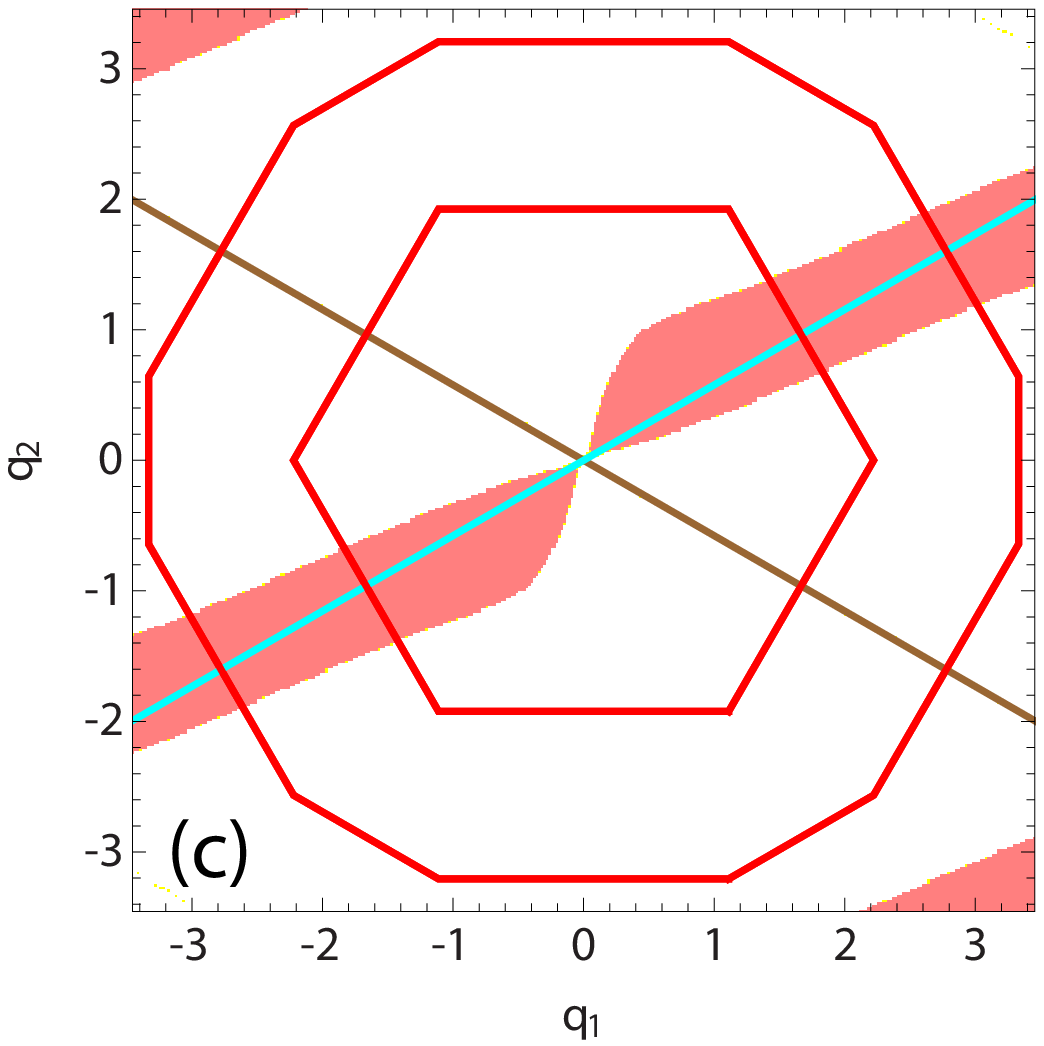}
\includegraphics[width=4.2cm]{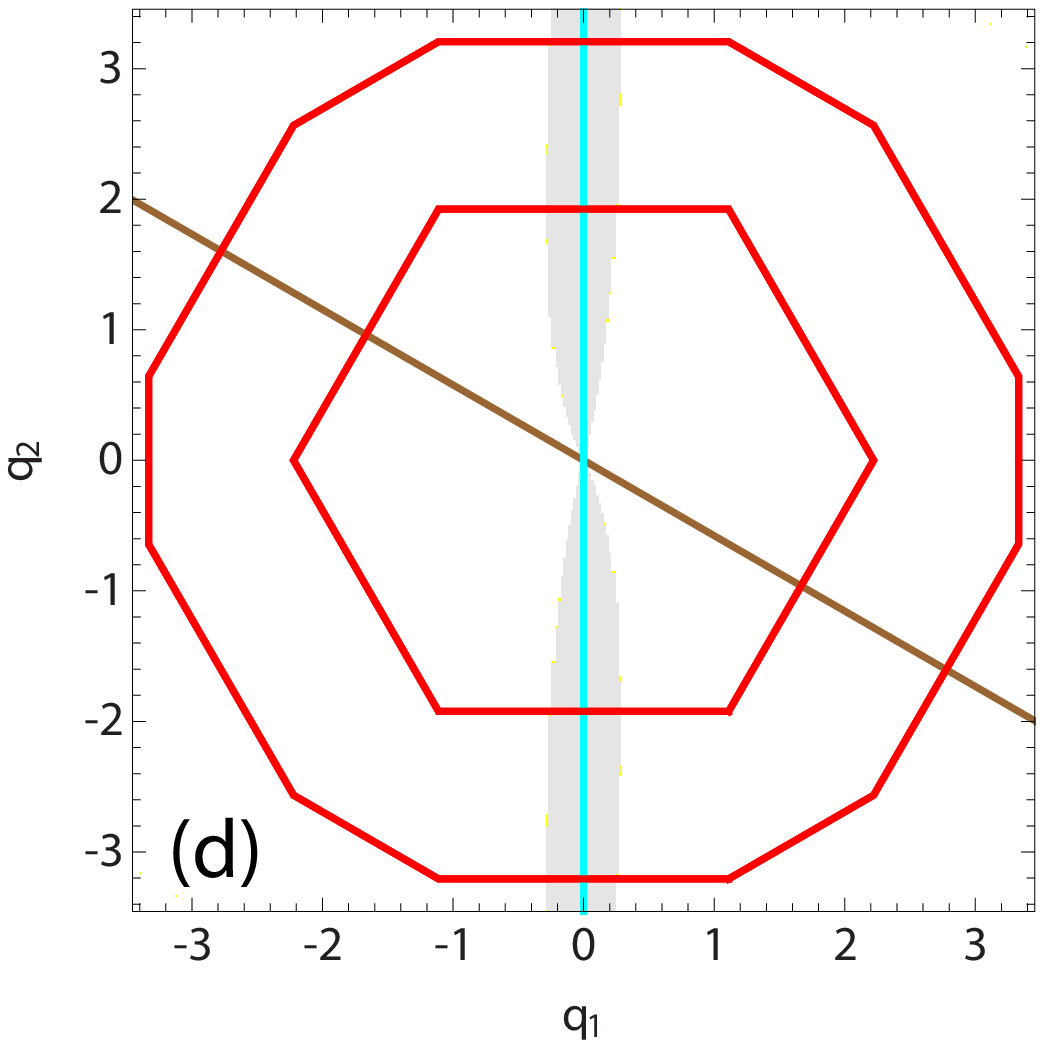}
\caption{(Color Online) Winding numbers $\wn_1$, $\wn_2$ and
$\wn_3$ for (a) $X_{1}$  and (b,c,d) $X_{-1}$. $q_1$ and $q_2$ are the components of
the respective $\qv_\perp$s. The (red) polygons depict
projections of the edges of the top and equatorial surfaces of the BZ onto these planes.
The black polygon in (a) represents the surface BZ. The
color-coding of the shaded
areas is as follows: pink corresponds to winding number 1,
white to 0, gray to -1, and black to -2. Yellow indicates
points where the numerical integration failed to converge properly.
The (brown) solid lines indicate the 2-fold degenerate zero
mode along the $(1, 1, 1)$-direction. The (cyan) light line
indicates the $(1,1, -1)$-direction and is a guide to the eye.}
\label{fig:weylPlots}%
\end{figure}

The topological properties of the phononic band structure of
the GPL are determined by the $\Cv$ or $\Qv$
matrixes~\cite{KaneLub2014}. The determinants of these matrices
map a path in $\brm{q}$-space onto a path in the complex plane.
Because $\brm{Q}$ and $\brm{C}$ are invariant under $\brm{q}
\to \brm{q} + \brm{G}$ for any reciprocal lattice vector
$\brm{G}$, any path in $\brm{q}$-space whose start and end
points are separated by a reciprocal lattice vector will map
onto a closed path in the complex plane. For simplicity, we
focus on paths in $\brm{q}$-space that are straight lines along
the primitive vectors $\bv_1$, $\bv_2$, $\bv_3$, satisfying
$\bv_i \cdot \Tv_j = 2 \pi \delta_{ij}$, of the reciprocal
lattice. The integer winding numbers of the corresponding
closed paths in the complex plane are
\begin{align}
\wn \left(\brm{q}_\perp, \Gv \right) = \frac{1}{2\pi i}
\int_{0}^{G} dp \, \frac{d}{d p} \, \mbox{Log} \mbox{Det} \,
\brm{Q} \left( \brm{q}_\perp, p, \Gv \right) ,
\end{align}
where $\brm{q}_\perp$ specifies the components of $\qv$ in the
surface BZ of the lattice plane defined by $\Gv$, $p$ is the
component of $\qv$ along $\Gv$, and $G = |\Gv|$. In fully
gapped systems, the winding numbers are independent of
$\qv_\perp$. In systems with Weyl singularities, they are not.
We will utilize this fact in the following to detect and map
out Weyl singularities in the GPL. The idea is to calculate the
winding numbers for an entire set of $\brm{q}_\perp$'s in a
given surface BZ. For example, for our integration along
$\bv_1$, we sweep the BZ that is spanned by the unit vectors
$\cvh_{1,1} = (\bv_1 \times\bv_2)/|\bv_1 \times \bv_2|$ and
$\cvh_{1,2} = (\bv_1 \times\cvh_{1,1})/|\bv_1 \times
\cvh_{1,1}|$. Figure~\ref{fig:weylPlots} compiles our results
for the winding numbers $\wn_i (\brm{q}_\perp) \equiv \wn
(\brm{q}_\perp,\Gv =\bv_i)$.
%%%%%%%%%%%%%%%%%%%%%%%%%%%
%%%%%%%%%%%%%%%%%%%%%%%%%%%
There is a qualitative difference between $X_{+1}$ and $X_{-1}$
in the values and distribution of their winding numbers: For
$X_{1}$, $\wn_i=-1$ throughout the surface BZ associated with
each $\bv_i$. For $X_{-1}$, boundaries determined by the
projections of Weyl lines onto any given surface BZ divide the
latter into regions with different winding numbers. In addition
to the Weyl lines, for both $X_1$ and $X_{-1}$ there is a
two-fold degenerate line of zero modes along the
$(1,1,1)$-direction (that we verify by diagonalizing $\brm{D}$
for $\brm{q}$ along $\nvh$) whose winding number is zero. This
line is a consequence of the $\qv_{\perp}= 0$ SSS of the flat
GKL planes in the GPL. In the more general version of our model
lattice, alluded to above, this line splits up into two
oppositely-charged Weyl lines when the reference positions of
one of the basis sites $1$, $2$ or $3$ is moved out of the GKL
plane. These Weyl lines are separated by a distance
proportional to the distance of the respective sites from the
GKL plane.

\begin{figure}[ptb]
%\vspace{-2.0 cm}
%\centering{\includegraphics[width=4.5cm]{weylLines3dWithCompPlot}}%
%\includegraphics[width=4.5cm]{weylLines3dWithCompPlot}
\includegraphics{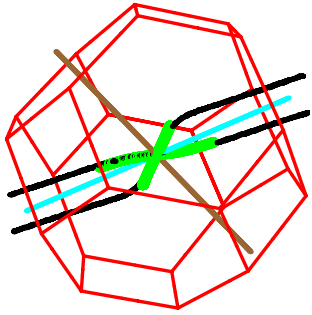}
\caption{(Color Online) Weyl lines for $X_{-1}$ traversing the
(red) dedocahedron-shaped BZ. The (black) points mark our
numerical reconstruction of the Weyl lines, the (green) lines
them stem from our analytical estimate based on the expansion
of $\det \brm{C}$, Eq.~(\ref{detCexpansion}). For $X_1$, only
the two-fold degenerate (brown) line along the
$(1,1,1)$-direction is present.}%
\label{fig:weylLines3dH1ZeroMinus1Plot}
\end{figure}

We can reconstruct the Weyl lines in $3d$ from their
projections onto the $2d$ BZ defined by the $\brm{b}_i$'s shown
in Fig~\ref{fig:weylPlots}. To this end, we use the following
approach: A point on a projected line with coordinates $(q_1,
q_2)$ in the plane perpendicular to $\bv_1$ is replaced by
$\brm{q} = p \bvh_1 + q_1 \cvh_{1,1} + q_2 \cvh_{1,2}$, where
$\bvh_1 = \brm{b}_1/| \brm{b}_1|$, and likewise for the
projection onto the plane perpendicular to $\brm{b}_2$. Then,
we calculate the intersections of the resulting manifolds.
Finally, we discard all intersection points whose projections
onto the plane perpendicular to $\brm{b}_3$ are incompatible
with our results for $\wn_3$. Our results obtained in this way
for the Weyl lines in 3$d$ are depicted in
Fig.~\ref{fig:weylLines3dH1ZeroMinus1Plot}. As follows from the
previous paragraph, the only zero modes that we encounter for
$X_1$ lie on a line along $\nvh$. For $X_{-1}$ there is in
addition a pair of Weyl lines. These Weyl lines do not form
closed loops but follow a path in the BZ between points
separated by a reciprocal lattice vector.

It is informative to compare our numerical results to
analytical predictions for the Weyl lines. To this end, we
consider small deviations about the original pyrochlore lattice
as parameterized by $X_\varepsilon \equiv (\varepsilon,
\varepsilon, \varepsilon, \varepsilon)$ and $X_{-\varepsilon}
\equiv (-\varepsilon, \varepsilon, \varepsilon, \varepsilon)$,
and we expand $\det \Cv(\qv)$ in powers of the $q_i$ and
$\varepsilon$. For both $X_\varepsilon$ and $X_{-\varepsilon}$,
the resulting expansion is of the form
\begin{align}
\det \brm{C} = f^{(3)} (\brm{q})\, \varepsilon^3 + f^{(4)}
(\brm{q})\, \varepsilon^2 + f^{(5)} (\brm{q})\, \varepsilon + O
(q_i^6) \, ,
\label{detCexpansion}
\end{align}
where the $f^{(m)} (\brm{q})$ are different functions of $m$th
order in $q_i$ for $X_{\varepsilon}$ and $X_{-\varepsilon}$
that vanish for any $\qv$ along $\nvh$. We solve for the
wave-vectors that are zeros of the right hand side of
Eq.~(\ref{detCexpansion}). For $X_\varepsilon$, there is only
one real solution that corresponds to any $\brm{q}$ along
$\nvh$. For $X_{-\varepsilon}$, on the other hand, there are
additional real solutions, which we display for $\varepsilon =
0.1$ in Fig.~\ref{fig:weylLines3dH1ZeroMinus1Plot}. Note the
nice agreement at small $q$ between the numerical and
analytical results for $X_1$ and $X_{-1}$.
%%%%%%%%%%%%%%%%%%%%%%%%%%%
%%%%%%%%%%%%%%%%%%%%%%%%%%%

Now, we turn to surface modes. In general, the total edge index
$\nu (\qv_\perp ,\Gv) = n_0(\qv_\perp ,\Gv) -
s(\qv_{\perp},\Gv)$ is the sum of a local part $\nu_L (\Gv)$
\cite{KaneLub2014,LubenskySun2015}, which is independent of
$\qv_{\perp}$, and a topological part $\nu_T(\qv_{\perp},\Gv
)$. At free surfaces, $s(\qv_{\perp},\Gv)=0$, and $\nu
(\qv_\perp,\Gv)=n_0(\qv_\perp,\Gv)$. The local count is
$\Gv\cdot \Rv_L/(2 \pi)$, where $\Gv$ is the \emph{outer}
normal to its lattice plane and $\Rv_L$ is the difference
between the dipole moment of the surface unit cell and that of
the reference cell of Fig.~\ref{fig:origAndGenLattice} (See
supplemenary material). In systems without Weyl singularities,
the winding numbers $\wn_i$ are independent of wave vector as
mentioned above and define a topological charge $\Rv_T =
\Remove{+} \sum_i \wn_i \Tv_i$. For $X_1$, in particular,
$\Rv_T = -(2,2,2)$. The topological surface count in these
systems is simply $\nu_T (\Gv)= \brm{G}\cdot \Rv_T/(2 \pi)$,
independent of $\qv_{\perp}$. In systems with Weyl points the
topological count $\nu_T (\qv_\perp,\Gv) = \wn
\left(\brm{q}_\perp, \Gv \right) $ depends on $\qv_\perp$ and
is not defined globally.

\begin{figure}[ptb]
\includegraphics{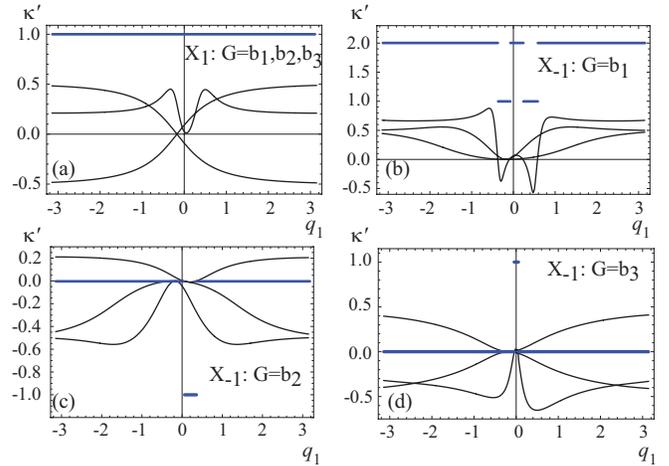}
\caption{(Color Online) $\kappa'$ (black lines) and $\nu_T$
(blue straight lines) of surface zero modes with fixed $q_2=0.1$ for
(a) $X_{1}$ and (b, c, d) $X_{-1}$.}
\label{fig:surfaceModes}%
\end{figure}

For simplicity, we focus here on surfaces whose normals are
parallel to primitive vectors $\bv_i$ of the reciprocal
lattice.  We calculate the complex inverse penetrations depths
$\kappa(\qv_{\perp})$ by setting $\brm{q} = i \kappa \bv_1 +
q_1 \cvh_{1,1} + q_2 \cvh_{1,2}$, and similarly for $i=2, 3$,
and solving for the roots of $\det \brm{C} (i\kappa, q_1, q_2)
= 0$.  Positive values of $\kappa' \equiv \text{Re} (\kappa)$
correspond to zero modes that decay in the direction of $\bv_1$
and that are, therefore, localized on the surface with outer
normal along $-\bv_1$.  Negative values of $\kappa'$ correspond
to states localized on the opposite surface with outer normal
$\bv_1$.  Figure~\ref{fig:surfaceModes} presents plots of
$\kappa'$ and $\nu_{T,i} = \nu_T (-\bv_i)$ as a function of
$q_1$ and fixed $q_2 = 0.1$ for our three surface orientations
and for $X_1$and $X_{-1}$.  For each $X$ and $\bv_i$, there are
positive and negative values of $\kappa'$ indicating
localization on both surfaces, but as required by the
Calladine-Maxwell theorem, there is always a total of three
zero modes on the two surfaces. For $X_1$, the three $\kappa'$s
are the same function of $q_1$ for all $\bv_i$s.  In addition
$\nu_{T,i}=1.0$ is independent of $i$. This implies that
$\nu_{L,i} =  \nu_{L} (-\bv_i) = 1.0$ for every $i$. $\nu_L$ is
a property of a surface that does not change if the topological
class is varied by changing $X$. For $X_{-1}$, the functions
$\kappa'(q_1)$ are different for the different surfaces, and
the number of positive and negative values undergo
discontinuous changes in accord with similar changes in
$\nu_{T,i} (q_1)$.  At the $-\bv_1$ surface, $\nu_{T1}$ takes
on values of $2$ and $1$, and, as required, the number of modes
localized on the $-\bv_1$ surface changes from $2$ to $1$ and
back again with the jumps in $\nu_{T1}$. Similarly, for
$-\bv_2$  surface, $\nu_{T,2}$ takes on values $0$ and $-1$ and
$n_0(q_1,-\bv_2)$ values $1$ and $0$ (though the latter region
is relatively small); and for the $-\bv_3$ surface, $\nu_{T3}$
is zero almost everywhere except for a very small region near
the origin where it is equal to $1$ and $\nu_3(q_1,-\bv_3)$ is
either $1$ or $0$. In the supplementary material, we provide
more detail about the calculation of $\Rv_L$, and we show
results for surfaces perpendicular to the $\Gv = \pm 2 \pi
(1,0,0)$ reciprocal lattice vector.  In this case, four bonds
must be cut to liberate a strip, and there are four surface
zero modes distributed between the two surfaces.
%%%%%%%%%%%%%%%%%%%%%%%%%%%
%%%%%%%%%%%%%%%%%%%%%%%%%%%

In conclusion, we have studied topological phonons in 3$d$ in a
generalized pyrochlore lattice. Our model lattice displays
distinct topological states, in the form of Weyl lines that
traverse the Brillouin Zone, and thereby underscores the
validity in 3$d$ of the general theory for topological phonons
laid out in Ref.~\cite{KaneLub2014}. Together with the recent
work on generalized square lattices~\cite{RocklinLub2016} and
2$d$ models of jammed matter~\cite{SussmanLub2016}, our work
hints that Weyl singularities are a common feature in Maxwell
lattices and that the GKL is special in that its unit cell does
not provide enough degrees of freedom to have them. The present
work indicates that the GPL is a candidate for detecting Weyl
lines in mechanical experiments. Given the advancement of 3$d$
printing technology, the creation of a GPL-like meta-material
in a lab should be within reach in the foreseeable future.

This work was supported by the NSF under No.~DMR-1104701 (OS,
TCL), No.~DMR-1120901 (OS, TCL). CLK was supported by a Simons
Investigators Grant and TCL by a Simons Fellows grant.

\newpage
\section{Supplemental material}
\subsection{Equilibrium matrix}
In terms of the primitive translation vectors, the position of
the origin (site 4) of a unit cell I can be labeled by 3
integers $l_1$, $l_2$, and $l_3$, $ \brm{T} (l_1, l_2, l_3) =
l_1 \brm{T}_1 + l_2 \brm{T}_2 + l_3 \brm{T}_3\ $. Using the
abbreviations $\brm{\tau}_1 = \brm{T} (-1, 1, 0)$,
$\brm{\tau}_2 = \brm{T} (0, -1, 1)$, $\brm{\tau}_3 = \brm{T}
(1, 0, -1)$, $\brm{\tau}_4 = \brm{T} (1, 0, 0)$, $\brm{\tau}_5
= \brm{T} (0, 1, 0)$ and $\brm{\tau}_6 = \brm{T} (0, 0, 1)$,
the equilibrium matrix of our model lattice in $\brm{q}$-space
reads
\begin{widetext}
\begin{align}
\brm{Q} (\brm{q}) = \left(
\begin{array}{cccccccccccc}
-\hat{a}_{1,1} & 0 & \hat{a}_{3,1} & \hat{a}_{4,1} & 0 & 0&
\hat{a}^\prime_{1,1} e^{-i \brm{q}\cdot \brm{\tau}_1}& 0 & -
\hat{a}^\prime_{3,1} & - \hat{a}^\prime_{4,1} & 0 & 0
\\
-\hat{a}_{1,2} & 0 & \hat{a}_{3,2} & \hat{a}_{4,2} & 0 & 0&
\hat{a}^\prime_{1,2} e^{-i \brm{q}\cdot \brm{\tau}_1}& 0 & -
\hat{a}^\prime_{3,2} & - \hat{a}^\prime_{4,2} & 0 & 0
\\
-\hat{a}_{1,3} & 0 & \hat{a}_{3,3} & \hat{a}_{4,3} & 0 & 0&
\hat{a}^\prime_{1,3} e^{-i \brm{q}\cdot \brm{\tau}_1}& 0 & -
\hat{a}^\prime_{3,3} & - \hat{a}^\prime_{4,3} & 0 & 0
\\
\hat{a}_{1,1} & - \hat{a}_{2,1} & 0& 0 & \hat{a}_{5,1} & 0& -
\hat{a}^\prime_{1,1} & \hat{a}^\prime_{2,1} e^{-i \brm{q}\cdot \brm{\tau}_2} &
0 & 0& - \hat{a}^\prime_{5,1} & 0
\\
\hat{a}_{1,2} & - \hat{a}_{2,2} & 0& 0 & \hat{a}_{5,2} & 0& -
\hat{a}^\prime_{1,2} & \hat{a}^\prime_{2,2} e^{-i \brm{q}\cdot \brm{\tau}_2} &
0 & 0& - \hat{a}^\prime_{5,2} & 0
\\
\hat{a}_{1,3} & - \hat{a}_{2,3} & 0& 0 & \hat{a}_{5,3} & 0& -
\hat{a}^\prime_{1,3} & \hat{a}^\prime_{2,3} e^{-i \brm{q}\cdot \brm{\tau}_2} &
0 & 0& - \hat{a}^\prime_{5,3} & 0
\\
0 & \hat{a}_{2,1} & - \hat{a}_{3,1} & 0& 0 & \hat{a}_{6,1} & 0
& -\hat{a}^\prime_{2,1} & \hat{a}^\prime_{3,1} e^{-i \brm{q}\cdot
\brm{\tau}_3} & 0 & 0 & - \hat{a}^\prime_{6,1}
\\
0 & \hat{a}_{2,2} & - \hat{a}_{3,2} & 0& 0 & \hat{a}_{6,2} & 0
& -\hat{a}^\prime_{2,2} & \hat{a}^\prime_{3,2} e^{-i \brm{q}\cdot
\brm{\tau}_3} & 0 & 0 & - \hat{a}^\prime_{6,2}
\\
0 & \hat{a}_{2,3} & - \hat{a}_{3,3} & 0& 0 & \hat{a}_{6,3} & 0
& -\hat{a}^\prime_{2,3} & \hat{a}^\prime_{3,3} e^{-i \brm{q}\cdot
\brm{\tau}_3} & 0 & 0 & - \hat{a}^\prime_{6,3}
\\
0 & 0 & 0 & - \hat{a}_{4,1} & - \hat{a}_{5,1} & - \hat{a}_{6,1}
& 0 & 0 & 0 & \hat{a}^\prime_{4,1} e^{-i \brm{q}\cdot\brm{\tau}_4} &
\hat{a}^\prime_{5,1} e^{-i \brm{q}\cdot \brm{\tau}_5} & \hat{a}^\prime_{6,1}
e^{i \brm{q}\cdot \brm{\tau}_6}
\\
0 & 0 & 0 & - \hat{a}_{4,2} & - \hat{a}_{5,2} & - \hat{a}_{6,2}
& 0 & 0 & 0 & \hat{a}^\prime_{4,2} e^{-i \brm{q}\cdot\brm{\tau}_4} &
\hat{a}^\prime_{5,2} e^{-i \brm{q}\cdot \brm{\tau}_5} & \hat{a}^\prime_{6,2}
e^{i \brm{q}\cdot \brm{\tau}_6}
\\
0 & 0 & 0 & - \hat{a}_{4,3} & - \hat{a}_{5,3} & - \hat{a}_{6,3}
& 0 & 0 & 0 & \hat{a}^\prime_{4,3} e^{-i \brm{q}\cdot\brm{\tau}_4} &
\hat{a}^\prime_{5,3} e^{-i \brm{q}\cdot \brm{\tau}_5} & \hat{a}^\prime_{6,3}
e^{i \brm{q}\cdot \brm{\tau}_6}
\end{array}
\right) \nonumber
\end{align}
\end{widetext}
where the $\hat{a}$'s and $\hat{a}^\prime$'s here are the
normalized interior and exterior bond vectors of the
generalized lattice.

\subsection{Surface with $\Gv = \bv_1$ -- unit cell and local count}
\begin{figure}[ptb]
\centering{\includegraphics[width=8.5cm]{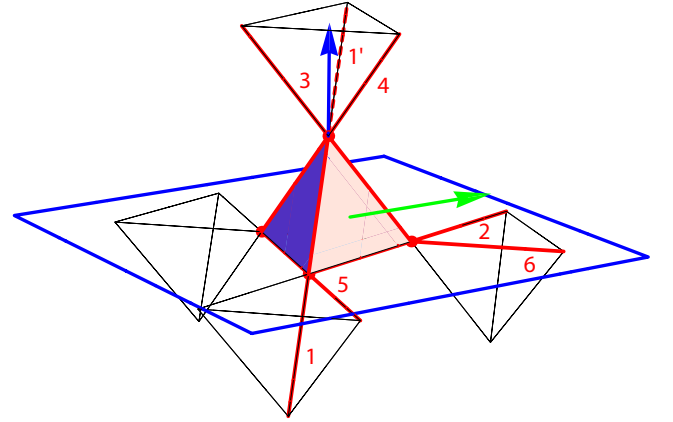}}
\caption{(Color Online) Our unit cell and a surface with inward normal along $\bv_1$.
The rectangular (blue) frame indicates the surface to be liberated and the
upward (blue) arrow its normal. The numbered red lines symbolize the
exterior bonds of the unit cell.}
\label{fig:origUintCellQ1}%
\end{figure}

Here, we provide some additional information about the
calculation of the surface zero modes for surfaces with normals
along the primitive reciprocal lattice vector $\Gv=\pm\bv_1$.
As shown in Fig.~\ref{fig:origUintCellQ1}, we have to cut 3
bonds ($1',3,4$ in the figure) per surface unit cell to
liberate  a slab with periodic boundary conditions
perpendicular to $\Gv$ and opposite surfaces parallel to the
plane defined by the (blue) quadrilateral.  At the lower
surface with $\bv_1$ the \emph{inner} normal, the exterior bond
$1$  protrudes, and at the upper surface with $\bv_1$ the
\emph{outer} normal, bonds $3$ and $4$ protrude.  Note that the
lower surface in this case is essentially flat whereas the
upper one has periodically spaced upward pointing tetrahedra.
To convert the reference unit cell to one compatible with the
lower surface with $\Gv= - \bv_1 = \pi (-1,-1,1)$, we have only
to move exterior bond $1$ by the vector $\Tv_b =
-(\av_1+\av_1^{\prime}) = \Tv_1 - \Tv_2 = (1,0,-1)$, which
implies that $\Rv_L^{\text{lower}}= 3 \sum_s \Delta \rv_s -
\sum_b \Delta \rv_b = -\Tv_1 + \Tv_2$, where $\Delta \rv_s$ and
$\Delta \rv_b$ are, respectively, the displacement of sites and
bonds from the reference unit cell to the surface compatible
one. Thus $\nu_L^{\text{lower}}= (-\bv_1/(2\pi))\cdot
\Rv_L^{\text{lower}}=1$.  To convert the reference unit cell to
one compatible with the upper surface, the protruding outer
bonds $3$ and $4$ must be displaced by $\Tv_3-\Tv_1$ and
$-\Tv_1$, respectively, yielding $\Rv_L^{\text{upper}}= 2 \Tv_1
- \Tv_3$ and
$\nu_L^{\text{upper}}=\bv_1\cdot\Rv_L^{\text{upper}}/2 \pi= 2$.
The topological polarization for the $X_1$ lattice is $\Rv_T =
-(2,2,2)$, so that $\nu_T^{\text{lower}}=-\bv_1\cdot\Rv_T = 1$
and $\Rv_T^{\text{upper}}= -1$.  Thus, $\nu^{\text{lower}}=
\nu_L^{\text{lower}}+\nu_T^{\text{lower}}=n_0(-\bv_1)=1+1 = 2$
and $\nu_T^{\text{upper}}=n_0(\bv_1)= 2-1=1$ in agreement with
Fig.~(\ref{fig:surfaceModes}).

%%%%%%%%%%%%%%%%%%%%%%%%-
%%%%%%%%%%%%%%%%%%%%%%%%%%%

\subsection{Surface with $\Gv \parallel (1,0,0)$}
To liberate a slab with surfaces perpendicular to $\Gv_x = 2
\pi (1,0,0)=\bv_1 + \bv_3$, we need to cut the four exterior
bonds $1'$, $2$, $4$, and $6$ in
Fig.~\ref{fig:surfaceModesG100}, and we expect total of $4$
zero modes as the displayed in Fig.~\ref{fig:surfaceModesG100}.
To convert the reference cell to one compatible with the lower
surface with $\Gv^{\text{lower}}=-\Gv_x$, we need, as in the
previous example, displace only exterior bond $1$, so that
$\Rv_L^{\text{lower}} = (-1,0,1)$ to produce
$\nu_L^{\text{lower}}= -\Gv_x \cdot \Rv_L^{\text{lower}}/2 \pi
= 2$.  To create a cell compatible with the upper surface, we
need to displace exterior bond $2$ by $-(\av_2+\av_2^{\prime})
= -\Tv_3 + \Tv_2$, bond $4$ by $ -\Tv_1$, and bond $6$ by $
-\Tv_3$.  These moves yield $\Rv_L^{\text{upper}} = 2 \Tv_3 -
\Tv_2 + \Tv_1$ and $\nu_T^{\text{upper}} = 3$, and
$\nu^{\text{upper}} = n_0(\Gv_x) = 3-2 = 1$ in agreement with
the three decaying modes (lower surface) and one growing mode
(upper surface) of Fig.~\ref{fig:surfaceModesG100}.
Similar analyses can be applied to the $X_1$ and other lattices
and to other surfaces and domain walls.

%%%%%%%%%%%%%%%%%%%%%%%%%%%
\begin{figure}[t]
	\centering{\includegraphics[width=3.25 in]{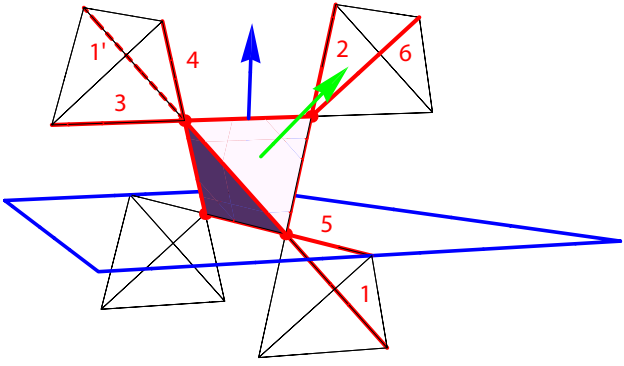}}
	\caption{(Color Online) Our unit cell and a surface with inward normal along $(1,0,0)$. The color-coding is as in Fig.~\ref{fig:origUintCellQ1}.}
	\label{fig:origUintCellG100}
\end{figure}
%%%%%%%%%%%%%%%%%%%%%%%%%%

%%%%%%%%%%%%%%%%%%%%%%%%%%%
\begin{figure}[tbp]
\includegraphics[width=3.2 in]{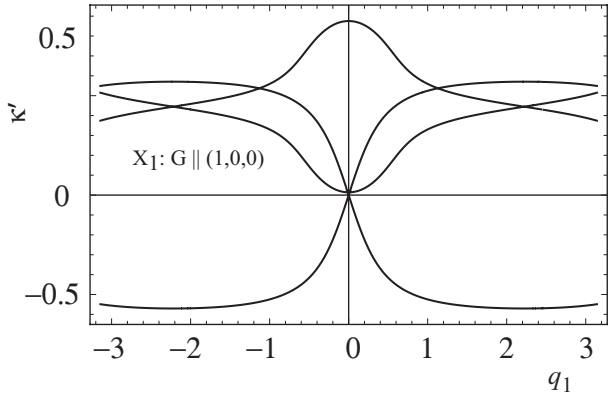}
\caption{(Color Online) $\kappa^\prime$ of surface zero modes for $\Gv = 2\pi (1,0,0)$ and fixed $q_2=0.1$ for $X_{1}$.}
\label{fig:surfaceModesG100}
\end{figure}
%%%%%%%%%%%%%%%%%%%%%%%%%%%

\end{document}